# Reference material for natural radionuclides in glass designed for underground experiments

P. P. Povinec[a*], M. K. Pham[b], J. Busto[c], C. Cerna[c,d], D. Degering[e], Y. Hamajima[f], K. Holý[a], M. Hult[g], M. Ješkovský[a], M. Köhler[e], A. Kováčik[a], M. Laubenstein[h], P. Loaiza[i], F. Mamedov[j], C. Marquet[d], J. Mott[k], M. Műllerová[a], F. Perrot[d], F. Piquemal[d,i], J.-L. Reyss[l], R. Saakyan[k], H. Simgen[m], B. Soulé[d], J. Staníček[a], I. Sýkora[a], I. Štekl[j]

[a] *Department of Nuclear Physics and Biophysics, Faculty of Mathematics, Physics and Informatics, Comenius University, SK-84248 Bratislava, Slovakia*
[b] *Environment Laboratories, International Atomic Energy Agency, MC 98000 Monaco*
[c] *Centre de Physique des Particules de Marseille, IN2P3/CNRS, F-13288 Marseille, France*
[d] *Centre Etudes Nucléaires de Bordeaux Gradignan, CNRS/IN2P3, F-33175 Gradignan, France*
[e] *Verein für Kernverfahrenstechnik und Analytik Rossendorf e.V., D-01314 Dresden, Germany*
[f] *Low-level Radioactivity Laboratory, University of Kanazawa, Wake, Nomi, 9231224 Japan*
[g] *Institute for Reference Materials and Measurements, Joint Research Centre, B-2440 Geel, Belgium*
[h] *Laboratori Nazionali del Gran Sasso, I-67100 Assergi (AQ), Italy*
[i] *Laboratoire Souterrain de Modane, F-73500 Modane, France*
[j] *Institute of Experimental and Applied Physics, Czech Technical University, CZ-12800 Prague, Czech Republic*
[k] *University College London, WC1E 6BT, United Kingdom*
[l] *Laboratoire des Sciences du Climat et de l'Environnement, CNRS, F-91198 Gif-sur-Yvette cedex, France*
[m] *Max-Planck-Institute für Kernphysik, D-69117 Heidelberg, Germany*

**Abstract** A reference material designed for the determination of natural radionuclides in solid samples (glass pellets) is described and the results of certification are presented. The material has been certified for 7 natural radionuclides ($^{40}$K, $^{226}$Ra, $^{228}$Ra, $^{228}$Th, $^{232}$Th, $^{235}$U and $^{238}$U). An information value is given for $^{210}$Pb. Radon ($^{222}$Rn) emanation experiments showed results comparable within participating laboratories, however, the number of data and precision was too low to carry out a certification process. The reference material may be used for quality management of analytical laboratories engaged in the high-sensitive analysis of radionuclides in the construction materials of detectors placed in ultra low background underground laboratories.

**Introduction**

The accurate and precise determination of radionuclide concentrations in construction materials of large detector assemblies operating in underground laboratories is essential for the assessment of the radiopurity of materials, necessary for reaching background levels prescribed for investigation of rare nuclear processes. Reference materials play a fundamental role in quality assessment of radionuclide analysis of construction materials reported by participating laboratories, especially in applications where ultra low-level activities need to be measured. They are valuable standards for method development and validation, and they can indicate the need to improve or change existing analytical methods [1-4].

Investigation of rare nuclear processes, such as double beta-decay experiments [5-8], neutrino interactions [9,10] and a search for dark matter [11-13], requires use in the detector environment of radioactively pure materials with minimum contamination by natural radionuclides. Except of U and Th (and their decay products), which can be found in almost all construction materials, other possible radionuclide contaminants could include $^{40}$K, which is frequently found in construction materials made of glass, e.g. in photomultipliers. Anthropogenic radionuclides, e.g. $^{137}$Cs of the Chernobyl and atmospheric atomic bomb tests origin [14], could be found in some materials exposed to the air, similarly as $^{60}$Co which is often found in steel samples. The acceptable contamination levels differ between various experiments. In some cases they should reach levels down to 1 nBq/g of material [15] when "zero" background detection systems have to be used for investigation of rare nuclear processes and decays [5-8].

---

[*] Corresponding author: e-mail: povinec@fmph.uniba.sk

Radiopurity of construction materials is thus a must if planned detection limits should be achievable. Usually several laboratories are engaged in radiopurity measurements of construction materials, therefore harmonization of measurements and validation of results is an important pre-requisite for producing data acceptable for further development of the detector systems. Reference materials, regularly analyzed by participating laboratories may thus represent a way how to reach these goals.

Radon isotopes ($^{222}$Rn – radon, $^{220}$Rn - thoron), represent a specific case of radionuclide contamination, which is usually the most dangerous for the operation of underground detectors because radon can diffuse through different parts of the detector structure [16]. Therefore it is important to know radon emanation characteristics of construction materials, as well as diffusion patterns of radon inside high sensitive detection systems, such as double beta-decay experiments, neutrino experiments, experiments on dark matter searches, etc.

The International Atomic Energy Agency (IAEA) has been organizing intercomparison exercises and developing reference materials within its AQCS (Analytical Quality Control Services) programme over three decades [1-4,17-19]. However, these reference materials have been mostly developed for environmental radioactivity studies, and their applications in radiopurity analysis of construction materials have been limited.

In this paper we describe a material based on glass pellets, designed for the determination of natural radionuclides, which could be used in laboratories participating in radiopurity analyses of construction materials. A trigger for this work has been requirements for quality control and interlaboratory checks of the laboratories responsible for radiopurity analyses of all materials used for the construction of the SuperNEMO detector. The SuperNEMO experiment will search for zero neutrino double beta-decay, and its first part (called Demonstrator) is presently under construction [4]. It will be operating from 2016 in the Modane underground laboratory (LSM). Heusser and Simgen (private communication and [20])) made a similar proposal in 2007 at the CELLAR (Collaboration of European Underground Laboratories) meeting to develop a solid-state natural activity standards for low-level counting, consisting of selected glass pellets, which have been analyzed for radionuclides at various laboratories.

To address the problem of data quality, and to assist collaboration laboratories in verifying their performance, the reference materials are important benchmarks in the quality management of laboratories.

**Materials and methods**

Description of the material

There are several requirements on the parameters and characteristics of reference materials. They should be chemically stable at least for 20 years, and the distribution of radionuclides in the material should be homogenous. Reasonable radionuclide levels in the material should be observed to reach good precision and accuracy of activity determination (e.g. $^{40}$K, $^{238}$U and $^{232}$Th decay products). Evacuation of the material should be possible, especially in the case when radon emanation experiments will be carried out.

The bunch of glass pellets was purchased in 2006, and after preliminary investigations (cleaning, homogeneity tests, preliminary activity measurements of gamma-emitters and radon emanation measurements) samples were distributed to SuperNEMO laboratories during 2007, and later also to expert laboratories participating in the intercomparison and in the certification process. Glass pellets (lead-free glass) of 2.1±0.1 mm in diameter with parameters listed in Table 1 were used in the experiment. The pellets were cleaned in an ultrasonic bath with detergent, then washed with ethyl alcohol, dried under inert gas, and finally dispatched to laboratories participating in the certification process. Before the measurements in a laboratory, the sample should be dried again and flushed with inert gas.

**Table 1** Characteristics of glass pellets used in radionuclide analyses

| | | |
|---|---|---|
| Chemical composition (%) | $SiO_2$ | 63 |
| | $Na_2O$ | 14 |
| | $CaO$ | 8 |
| | $Al_2O_3$ | 7 |
| | $B_2O_3$ | 5 |
| | $MgO$ | 3 |
| Specific weight (kg/m$^3$) | 2,500 ± 40 | |
| Coefficient of thermal extension ($10^{-8}$ K$^{-1}$) | 9.2 ± 0.4 | |
| Littleton softening point (°C) | 670 ± 10 | |
| Elasticity module (Mpa) | 7.75 | |
| Hydrolytic class | HGB 3 | |
| Acidic class (DIN 12116) | III | |
| Alkaline class (ISO 695) | A-1 | |
| Crushing strength (N) | 900 | |
| Number of pellets in 1 kg | 81,540 | |
| Number of pellets in 1 L | 128,180 | |
| Contact surface of 1 kg (cm$^2$) | 10,247 | |

Sample dispatch and data feedback

The test material was distributed to 10 participating laboratories, which represent well-known high precision laboratories working in low-level gamma-ray spectrometry, mostly with underground installations. The reference date for reporting activities was set at 1 August 2010. Participating laboratories were requested to carry out several independent analyses of glass pellets, and to determine in the sample as many radionuclides as possible. For each radionuclide analysed, the following information was requested from participating laboratories:
- (i) average weight of sample used for analysis;
- (ii) number of activity measurements per sample;
- (iii) resulting massic activity (Bq/kg) corrected for blank, background, etc.;
- (iv) estimation of the combined uncertainties;
- (v) description of cleaning procedures;
- (vi) description of the counting equipment;
- (vii) counting time and decay corrections.

Simultaneously, six laboratories were engaged in radon emanation experiments. The list of laboratories participating in the gamma-ray spectrometry and radon emanation experiments, and their abbreviations are listed in Table 2.

**Table 2** Laboratories participating in analysis of radionuclides and radon emanation experiments in glass pellets

| Name | Abbreviation | Analyses |
| --- | --- | --- |
| Comenius University, Bratislava, Slovakia | UNIBA | Radionuclides/Radon emanation |
| Max-Planck-Institute für Kernphysik, Heidelberg, Germany | MPI-KK | Radionuclides/Radon emanation |
| Laboratoire des Sciences du Climat et de l´Environnement, Gif-sur-Yvette, France | LSCE | Radionuclides |
| Institute of Reference Materials and Measurement, JRC, Geel, Belgium | IRMM | Radionuclides |
| International Atomic Energy Agency, Environment Laboratories, Monaco | IAEA | Radionuclides |
| Gran Sasso National Laboratory, Assergi, Italy | LNGS | Radionuclides |
| Verein für Kernverfahrenstechnik und Analytik Rossendorf e.V., Dresden, Germany | VKTA | Radionuclides |
| Low-level Radioactivity Laboratory, Kanazawa University, Japan | LLRL | Radionuclides |
| Laboratoire Souterrain de Modane, France | LSM | Radionuclides |
| Centre Etudes Nucléaires de Bordeaux Gradignan, France | CENBG | Radionuclides/Radon emanation |
| University College London, United Kingdom | UCL | Radon emanation |
| Centre de Physique des Particules de Marseille, France | CPPM | Radon emanation |
| Czech Technical University, Prague, Czech Republic | CTU | Radon emanation |

Data treatment

The massic activities of natural radionuclides measured in the glass sample were reported by participating laboratories. We have been following the statistical evaluation procedures developed in the International Atomic Energy Agency´s Environment Laboratories in Monaco [17,18]. Calculations were based on the assumption of non- parametric distribution of data to which distribution-free statistics are applicable. When not provided by the laboratory, the means were calculated from individual results either as arithmetic means with corresponding uncertainties when more than two similar results were reported, or as weighted means with weighted uncertainties in the case of only two results reported. The reporting values were checked for the presence of outliers using the Box and Whisker plot test. No outliers were identified in the data set formed from the laboratory reports. Medians were calculated from the results passing the test, rounded off to the most significant figure of the uncertainty. These values were considered to be the most reliable estimates of the true values. Confidence intervals were determined from a non-parametric sample population, and were expressed as two-sided intervals representing 95% confidence limits [17,18].

Following the International Union of Pure and Applied Chemistry [21] and International Organization for Standardization (ISO) [22] recommendations for assessment of laboratory performance, the z-score methodology was used in the evaluation of results. The z-score was calculated according to the formula

$$z = (x_i - x_a)/s_b$$

where $x_i$ is the robust mean of the massic activity values reported by laboratory $i$, $x_a$ is the assigned value (mean value of accepted results), and $s_b$ is the target standard deviation. The selection of the right target value depends on the objectives of the exercise. For radionuclide analysis, laboratories were required to have a relative bias below 20% ($s_b$<10%). The uncertainty of the assigned value ($s_a$) was included in the target value for bias using the equation [23]:

$$z = (x_i - x_a)/\sqrt{(s_b^2 + s_a^2)}.$$

The performance of laboratories in terms of accuracy was expressed by the z-scores for each radionuclide. The performance was considered to be acceptable if the difference between the robust mean of the laboratory and the assigned value (in $s_b$ units) was less than or equal to two. A z-score from 2 to 3 indicates that the results are of questionable quality. If $|z|>3$, the analysis would be considered to be out of control.

Criteria for certification

The certification process was carried out following the ISO Guide 35 [24] using the data reported by the participating laboratories. For data sets comprising 5 or more accepted laboratory means, median values and confidence intervals were calculated as estimations of true massic activities representing the best estimator of the property values. The calculated median activities were considered as the "certified value" when [19]:
   (i) at least 5 average values were available, calculated from at least 3 different laboratories, and
   (ii) the relative uncertainty of the median did not exceed 10% for activities from 10 to 100 Bq/kg, and 30% for activities lower than 10 Bq/kg.
An activity value was considered as an "information value" when at least 5 average values of the same order of magnitude calculated from the results of at least 2 different laboratories were available.

**Results and discussion**

Homogeneity tests

The homogeneity of the sample was checked by measuring $^{40}$K and $^{226}$Ra activities in 20 glass pellet samples taken at random. Non-destructive gamma-ray spectrometry was performed on 20 and 100 g samples. Homogeneity was determined using one-way analysis of variance. The coefficient of variation was below 10% for all radionuclides determined by gamma-ray spectrometry methods. The "between samples" variances of 100 g samples showed no significant differences from the "within sample" variances of 20 g samples for two radionuclides tested. Thus, the material was considered to be homogeneous for the tested radionuclides at the weight ranges used.

The z-score analysis

The performance of laboratories in terms of accuracy was tested by the z-score analysis for each measured radionuclide. The performance was considered to be acceptable if the z-score was ≤ 2. The z-score distributions for all laboratories and radionuclides were symmetric and with values below 2, indicating that the performance of the laboratories was satisfactory. A typical example of the z-score analysis for $^{226}$Ra is shown in Fig. 1. The z-score evaluation represents a simple method, which gives participating laboratories a normalized performance score for bias.

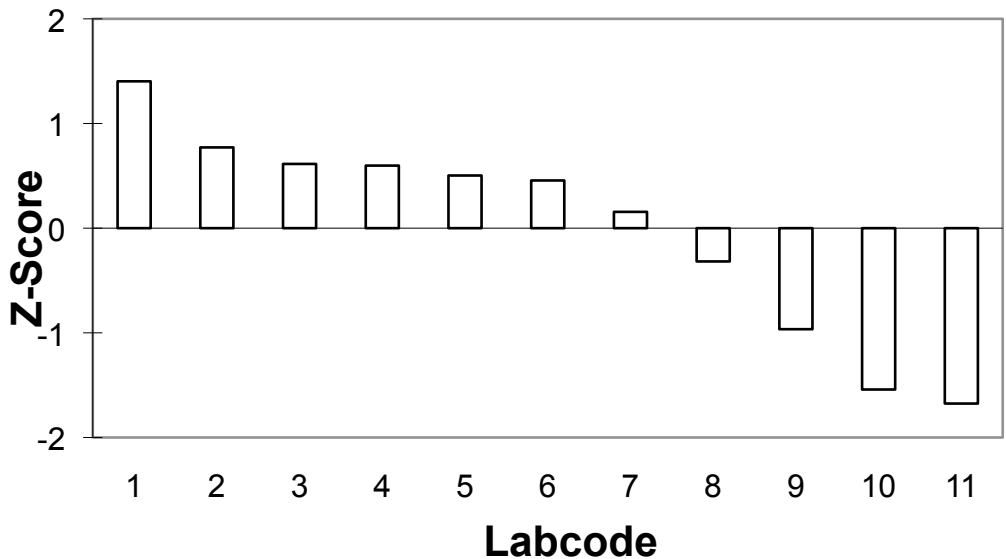

Fig. 1. The z-score analysis of $^{226}$Ra data.

Radionuclides with certified values

The massic activities of 7 natural radionuclides were reported by participating laboratories, which could gain the status of certified values ($^{40}$K, $^{226}$Ra, $^{228}$Ra, $^{228}$Th, $^{232}$Th, $^{235}$U and $^{238}$U). All results were obtained by nondestructive low-level gamma-ray spectrometry. As an example, Figs. 2 and 3 present the evaluation results in order of ascending massic activities for $^{226}$Ra and $^{238}$U. Also shown are the distribution medians and corresponding confidence intervals.

*$^{40}$K*

Eleven laboratory means obtained by gamma-ray spectrometry were available for data evaluation. It looks like that some of the laboratories encountered problems with calibration and the correct estimation of the background under the $^{40}$K full energy photopeak, which is usually present in the background gamma-ray spectra. The data showed reasonable homogeneity. The z-score values of accepted data were below 1.3, showing good performances by the laboratories. The median, given as the certified value, is 9.2 Bq/kg. The 95% confidence interval is (7.8–10.6) Bq/kg).

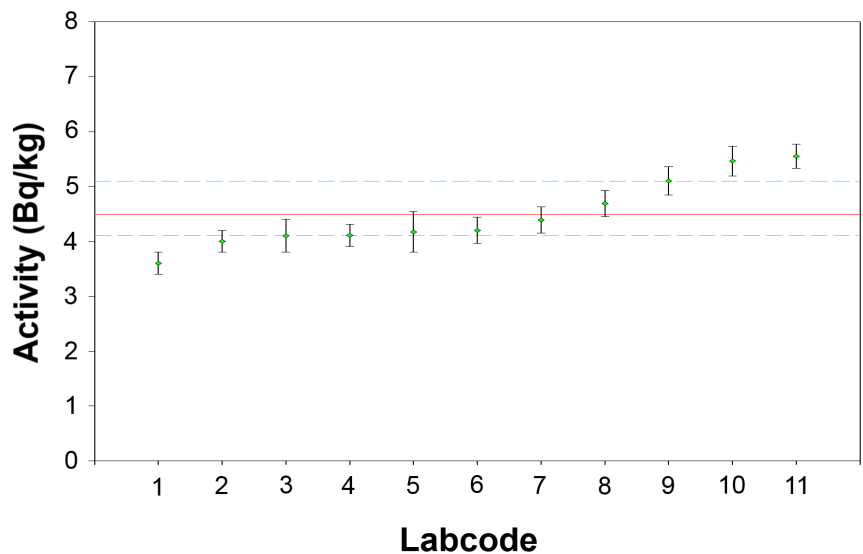

Fig. 2. Data evaluation for $^{226}$Ra in glass pellets (median and 95% confidence intervals are also shown). The error bars correspond to the standard combined uncertainties (k=1) reported by the laboratories.

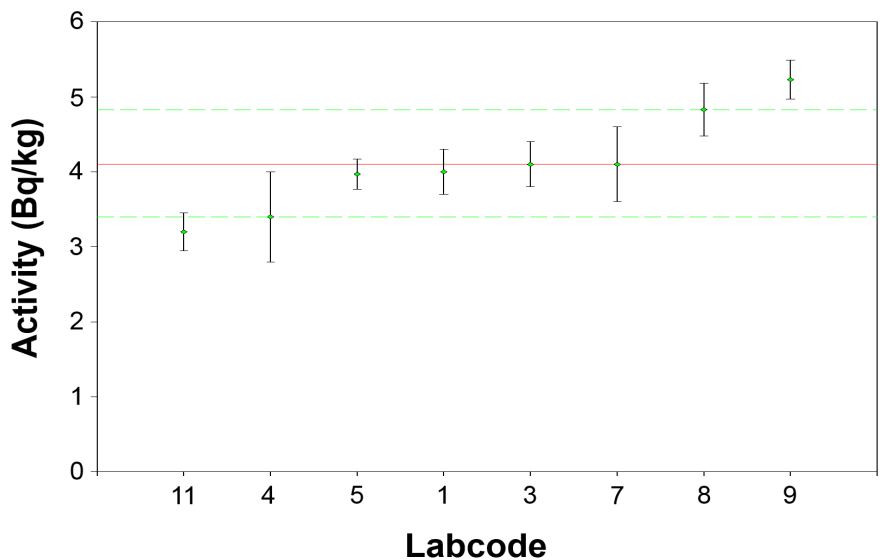

Fig. 3. Data evaluation for $^{238}$U in glass pellets (median and 95% confidence intervals are also shown). The error bars correspond to the standard combined uncertainties (k=1) reported by the laboratories.

## $^{226}$Ra ($^{214}$Bi, $^{214}$Pb)

Eleven laboratory means obtained by gamma-ray spectrometry were used in the evaluation (Fig. 2). The laboratories used for the activity calculations corresponding gamma-ray peaks of $^{226}$Ra, $^{214}$Bi and $^{214}$Pb. The efficiency calibration of HPGe spectrometers was carried out using either Monte Carlo codes or radioactivity (RA) standards. The z-score values were below 1.7. The median gives the certified value of 4.2 Bq/kg. The 95% confidence interval is (4.1–5.1) Bq/kg. Table 3 summarizes the data obtained for $^{226}$Ra. The uncertainties listed in Table 3 represent combined statistical and systematic uncertainties (e.g. from efficiency calibration, background estimation, etc.). Similar data sheets were prepared for all radionuclides.

**Table 3** Results of $^{226}$Ra measurements in glass pellets

| Lab code | Method – HPGe spectrometry | Number of results | Efficiency calibration | Mass (g) | $^{226}$Ra (Bq/kg) |
|---|---|---|---|---|---|
| 1 | $^{226}$Ra, $^{214}$Bi, $^{214}$Pb | 3 | Monte Carlo | 200 | 3.65 ± 0.22 |
| 2 | $^{226}$Ra, $^{214}$Bi, $^{214}$Pb | 1 | Monte Carlo | 100 | 4.0 ± 0.2 |
| 3 | $^{226}$Ra, $^{214}$Bi, $^{214}$Pb | 1 | Monte Carlo | 11.6 | 4.1 ± 0.3 |
| 4 | $^{226}$Ra, $^{214}$Bi, $^{214}$Pb | 3 | Monte Carlo | 211 | 4.11 ± 0.10 |
| 5 | $^{226}$Ra | 1 | RA standard | 150 | 4.17 ± 0.37 |
| 6 | $^{226}$Ra, $^{214}$Bi, $^{214}$Pb | 2 | Monte Carlo | 257 | 4.2 ± 0.2 |
| 7 | $^{226}$Ra, $^{214}$Bi, $^{214}$Pb | 1 | Monte Carlo | 100 | 4.39 ± 0.24 |
| 8 | $^{226}$Ra, $^{214}$Bi, $^{214}$Pb | 1 | RA standard | 5 | 4.69 ± 0.24 |
| 9 | $^{226}$Ra, $^{214}$Bi, $^{214}$Pb | 1 | Monte Carlo | 102 | 5.10 ± 0.26 |
| 10 | $^{214}$Bi, $^{214}$Pb | 2 | RA standard | 150 | 5.47 ± 0.27 |
| 11 | $^{226}$Ra, $^{214}$Bi, $^{214}$Pb | 3 | RA standard | 7.5-171 | 5.55 ± 0.22 |

| | |
|---|---|
| Number of reported lab. means | 11 |
| Number of accepted lab. means | 11 |
| Mean | 4.5 |
| Median | 4.2 |
| Confidence interval (α = 0.05) | 4.1-5.1 |

*Reported uncertainties are at 1 sigma level

### $^{228}$Ra ($^{228}$Ac)

Eight laboratory means obtained by gamma-ray spectrometry were used in the evaluation. The laboratories used for the activity calculations corresponding gamma-ray peaks of $^{228}$Ra and $^{228}$Ac. The z-score values were below 2.0. The median, given as the certified value is 2.4 Bq/kg. The 95% confidence interval is (2.1–2.6) Bq/kg.

### $^{228}$Th ($^{212}$Pb, $^{208}$Tl)

Eight laboratory means obtained by gamma-ray spectrometry were used in the evaluation. The laboratories used for the activity calculations corresponding gamma-ray peaks of $^{212}$Pb and $^{208}$Tl. The z-score values were below 1.5. The median, given as the certified value is 2.4 Bq/kg. The 95% confidence interval is (2.1–2.5) Bq/kg.

### $^{232}$Th ($^{228}$Ra, $^{228}$Th)

Eight laboratory means obtained by gamma-ray spectrometry were used in the evaluation. The laboratories used for the activity calculations corresponding gamma-ray peaks of $^{228}$Ra ($^{228}$Ac) and $^{228}$Th ($^{212}$Pb and $^{208}$Tl). The z-score values were below 1.5. The median, given as the certified value is 2.4 Bq/kg. The 95% confidence interval is (2.1–2.5) Bq/kg.

### $^{235}$U

Five laboratory means obtained by gamma-ray spectrometry were used in the evaluation. The z-score values were below 1.4. The median, given as the certified value is 0.21 Bq/kg. The 95% confidence interval is (0.19–0.25) Bq/kg.

### $^{238}U$ ($^{234}Th$, $^{234m}Pa$)

Nine laboratory means obtained by gamma-ray spectrometry were used in the evaluation (Fig. 3). The laboratories used for the activity calculations corresponding gamma-ray peaks of $^{234}Th$ and $^{234m}Pa$. The z-score values were below 1.3. The median, given as the certified value is 4.0 Bq/kg. The 95% confidence interval is (3.4–4.8) Bq/kg. The obtained results on certified radionuclides are summarized in Table 4.

**Table 4** Certified values for radionuclides in the reference material of glass pellets (the reference date: 1 August 2010)

| Radionuclide | Mean (Bq/kg) | Median (Bq/kg) | Confidence interval ($\alpha = 0.05$) (Bq/kg) | Number of Mean values* |
|---|---|---|---|---|
| $^{40}K$ | 8.8 | 9.2 | 7.8 – 10.6 | 11 |
| $^{226}Ra$ | 4.5 | 4.2 | 4.1 -5.1 | 11 |
| $^{228}Ra$ | 2.4 | 2.4 | 2.1 – 2.6 | 8 |
| $^{228}Th$ | 2.4 | 2.4 | 2.1 – 2.5 | 8 |
| $^{232}Th$ | 2.4 | 2.4 | 2.1-2.5 | 8 |
| $^{235}U$ | 0.21 | 0.21 | 0.19 – 0.25 | 5 |
| $^{238}U$ | 4.10 | 4.0 | 3.4 – 4.8 | 9 |

\* Number of accepted laboratory means which were used to calculate the certified activities and the corresponding confidence intervals

Radionuclides with information values

### $^{210}Pb$

Five laboratory means obtained by gamma-ray spectrometry were used in the evaluation. The laboratories used for the activity calculations the corresponding gamma-ray peak of $^{210}Pb$. It looks like that some of the laboratories encountered problems with calibration at lower energies of gamma-ray spectra where the $^{210}Pb$ peak is found (46 keV). The z-score values were below 1.6. The median, given as the information value is 3.9 Bq/kg. The 95% confidence interval is (2.4–5.2) Bq/kg. Summary of data for $^{210}Pb$ is presented in Table 5.

**Table 5** Information values for $^{210}Pb$ in the reference material of glass pellets (the reference date: 1 August 2010)

| Radionuclide | Mean (Bq/kg) | Median (Bq/kg) | Confidence interval ($\alpha = 0.05$) (Bq/kg) | Number of Mean values* |
|---|---|---|---|---|
| $^{210}Pb$ | 3.8 | 3.9 | 2.4 – 5.2 | 5 |

\* Number of accepted laboratory means which were used to calculate the information value and the corresponding confidence intervals

Less-frequently reported radionuclides

For $^{230}$Th only detection limit (< 4.4 Bq/kg) was reported by one laboratory. No anthropogenic radionuclides were detected in the samples of glass pellets. The corresponding detection limits were < 0.02 Bq/kg for $^{137}$Cs and 0.04 Bq/kg for $^{60}$Co.

Activity ratios

All data presented for $^{232}$Th and $^{238}$U and their daughters ($^{228}$Ra, $^{228}$Ac, $^{228}$Th, $^{212}$Pb and $^{208}$Tl for $^{232}$Th, and $^{234}$Th, $^{234m}Pa$, $^{226}$Ra, $^{214}$Bi, $^{214}$Pb and $^{210}$Pb for $^{238}$U) indicate that both decay series have been within statistical uncertainties in equilibrium. The measured activity ratio of $^{238}$U/$^{235}$U, equal to 19.0, is close to its nominal value of 21.5 observed in materials with natural isotopic abundance [25].

Radon emanation results

Radon ($^{222}$Rn) emanation experiments were carried out in six expert laboratories. Radioanalytical techniques used during experiments consisted of a radon emanation chamber, transfer of radon by a carrier gas to a calibrated volume and then to a detector, and finally radon counting by a solid-state detector, an ionization chamber or a scintillation detector. The obtained results are presented in Table 6. Some of the laboratories performed duplicate analyses with larger samples of glass pellets with the aim to improve their detection limits. Unfortunately, there is large scattering of data and 3 of 6 laboratories reported only detection limits. The statistical evaluation of the data for the certification purposes could not be therefore carried out. Although the spread of the results reported in Table 6 is large, they are still within 3 sigma standard deviations. The reasons behind such large spread of the results may be in radon adsorption on glass pellets and its removal in the detectors, as well as on the dependence of electrostatic properties of pellets on the local environment (e.g. during transfer of pellets after cleaning to the detector). The reported uncertainties represent, however, in this case only statistical counting uncertainties. As the total uncertainties (including systematic uncertainties) are expected to be by a factor of two larger, the reported results would be then within 2 sigma standard deviations. The weighted mean of saturated $^{222}$Rn activity in glass pellets calculated from results of 3 laboratories is 0.62± 0.08 mBq/kg. By comparing this result with data presented in Table 6 we see that the fraction of $^{222}$Rn emanated from $^{226}$Ra present in glass pellets is about 25 %.

of radon emanation measurements (uncertainties reported at 1 sigma)

| Parameter | Laboratory | | | | | |
| --- | --- | --- | --- | --- | --- | --- |
| | UNIBA | CENBG | CPPM | UCL | MPI-K | CTU |
| Mass (g) | 1355 | 426 | 166 | 125 | 238 | 356 |
| Surface (m$^2$) | 1.553 | 0.488 | 0.190 | 0.143 | 0.273 | 0.408 |
| Rn emanation (μBq/kg/s) | < 2.5 | 5.2 ± 1.7 | 3.5 ± 0.4 | <4.6 | 0.96 ± 0.15 | <1.3 |
| Rn emanation (μBq/m$^2$/s) | < 1.7 | 3.1 ± 0.5 | 2.9 ± 0.3 | <4.0 | 0.84 ± 0.13 | <1.1 |
| Rn activity* (mBq/kg) | < 1.3 | 2.5 ± 0.7 | 1.7 ± 0.3 | <2.2 | 0.46 ± 0.07 | <0.6 |
| Rn activity* weighted mean (mBq/kg) | 0.62± 0.08 | | | | | |

* The $^{222}$Rn activity in equilibrium with the $^{226}$Ra source

**Conclusions**

Reference material of glass pellets was described in this paper, and data on natural radionuclides were evaluated. The medians of massic activities with 95% confidence intervals were chosen as the most reliable estimates of the true values. Ten well-known expert laboratories took part in the gamma-spectrometry analyses of radionuclides. The material was certified as a reference material for 7 radionuclides ($^{40}$K, $^{226}$Ra, $^{228}$Ra, $^{228}$Th, $^{232}$Th, $^{235}$U, and $^{238}$U), and the information value was estimated for $^{210}$Pb. The reference date for decay corrections has been set at 1 August 2010.

Radon ($^{222}$Rn) emanation experiments performed by six laboratories showed results comparable within statistical uncertainties, however, three laboratories reported only detection limits, therefore the certification process was not carried out. The fraction of $^{222}$Rn emanated from $^{226}$Ra present in glass pellets was about 25 %.

This reference material may be used for quality management of analytical laboratories, assessing the validity of analytical methods and identifying weaknesses in methodologies. The reference material may be of special interest to laboratories engaged in the high-sensitive analysis of radionuclides in the construction materials of detectors placed in ultra low background underground laboratories. It is recommended to clean the glass pellets before analysis in an ultrasonic bath with detergent, then to wash them with ethyl alcohol, and finally to dry them in an oven. The amount available for applications in external laboratories is about 10 kg. In the case of interest to get a sample of the reference material please contact P. Povinec (povinec@fmph.uniba.sk).

**Acknowledgements** The authors are indebted to their numerous colleagues who took part in the analytical works. The Bratislava group acknowledges a partial support provided by the VEGA grant # 1/0783/14 from The Ministry of Education, Science, Research and Sport of the Slovak Republic, and by the EU Research and Development Operational Program funded by the ERDF (projects # 26240120012, 26240120026 and 26240220004). The IAEA is grateful for the support provided to its Environment Laboratories by the Government of the Principality of Monaco.